\documentclass{aa}  

\usepackage{graphicx}
\usepackage{txfonts}
\usepackage[colorlinks=true,citecolor=blue]{hyperref}
\usepackage{natbib}

\begin{document}

   \title{The Spatial Distribution of $\rm CH_4$ and $\rm CO_2$ Ice around Protostars IRAS 16253-2429 and IRAS 23385+6053}


   \author{Lei Lei
          \inst{1,2}
          \and
          Lei Feng
          \inst{1,2,3}
          \and 
          Yi-Zhong Fan
          \inst{1,2}
          }

   \institute{Key Laboratory of Dark Matter and Space Astronomy, Purple Mountain Observatory, Chinese Academy of Sciences, Nanjing 210023, China\\
              \email{fenglei@pmo.ac.cn; yzfan@pmo.ac.cn}
         \and
             School of Astronomy and Space Science, University of Science and Technology of China, Hefei 230026, China
             \and
             Joint Center for Particle, Nuclear Physics and Cosmology,
Nanjing University – Purple Mountain Observatory, Nanjing 210093, China\\
             }

   \date{Received x xx, xxxx; accepted x xx, xxxx}

 
  \abstract
   {The origin and evolution of organic molecules represent a pivotal issue in the fields of astrobiology and astrochemistry, potentially shedding light on the origins of life. The James Webb Space Telescope (JWST), with its exceptional sensitivity and spectral resolution, is well suitable to observe molecules such as methane ($\rm CH_4$).}
   {Our analysis focused on the distribution of $\rm CH_4$, $\rm CO_2$, $\rm H_2O$, $\rm{CH_3OH+NH_4^+}$ ice and silicate absorption dips at approximately 7.7, 15.0, 6.0, 6.7 and 10.0 micrometers in two protostars: IRAS 16253-2429 and IRAS 23385+6053.}
   {The two protostars IRAS 16253-2429 and IRAS 23385+6053 were observed by JWST Mid-Infrared Instrument (MIRI) Integral Field Unit (IFU) in 2022, which provided spectral information with wide wavelength range covered $\rm CH_4$, $\rm CO_2$, $\rm H_2O$, $\rm{CH_3OH+NH_4^+}$ ice and silicate absorption dips in infrared. We empoly the equivalent width (EW) to calculate the intensity of the absorption dips and emission lines in the JWST MIRI/IFU data. We extract the $\rm CH_4$, $\rm CO_2$, $\rm H_2O$, $\rm{CH_3OH+NH_4^+}$ ice EW maps and silicate extinction maps of the two sources.   }
   {Our result reveal that the spatial distribution of methane (\(\rm CH_4\)) in the protostellar system IRAS 16253-2429 closely mirrors that of its carbon dioxide (\(\rm CO_2\)) ice, forming a surrounded distribution that encircles the central protostar. This alignment suggests a common formation mechanism and subsequent trapping within the protostellar envelope, which is consistent with the "Classical" dark-cloud chemistry with ion-molecule reaction. In contrast, the spatial distributions of various molecules in the system IRAS 23385+6053 exhibit low similarities, which may be attributed to the dynamic influences of outflows or accretion processes. These discrepancies highlight the complex interplay between physical processes and chemical evolution in protostellar environments. } 
   {}


   \maketitle
%

\section{Introduction}

The formation, evolution, survival, transport, and transformation of organic molecules such as methane (\(\rm{CH_4}\)) and inorganic molecules like carbon dioxide (\(\rm{CO_2}\)) and water (\(\rm{H_2O}\)) are fundamental processes that are crucial for unraveling the mysteries of astrobiological \citep{Ehrenfreund:2000ke} and astrochemical evolution \citep{1973ApJ...185..505H,2012A&ARv..20...56C,2018IAUS..332....3V}. Historical theories have suggested that simple molecules and dust are synthesized during the star formation and evolution processes, and these constituents can further react under the influence of cosmic ray heating to form more complex molecules \citep{1973ApJ...185..505H,2005IAUS..231.....L,2012A&ARv..20...56C}.
Observations of the ices of these simple molecules have been facilitated by space-based infrared telescopes, including the Infrared Space Observatory (ISO) \citep{2004ApJS..151...35G}, the Spitzer Space Telescope \citep{2008ApJ...678..985B}, AKARI \citep{2012A&A...538A..57A}, the Stratospheric Observatory for Infrared Astronomy (SOFIA) \citep{2023ApJ...953..103L}, and the Herschel Space Observatory \citep{2012A&A...539A.132C,2021A&A...648A..24V}. However, studying the spatial distribution of molecular ices has been challenging due to limitations in the design of their spectrometers, which has historically constrained the detailed analysis of these distributions.

The JWST Mid-Infrared Instrument (MIRI, \cite{2015PASP..127..584R,2015PASP..127..595W,2023PASP..135f8001G}) Integral Field Unit (IFU) and the Near Infrared Spectrograph IFU (NIRSpec/IFU, \cite{2023PASP..135f8001G}) are capable of capturing the spectra of extended structures with a single exposure. Spectra can be extracted from each pixel within the IFU data, and images of the object at each wavelength can be constructed. This observational technique enables us to map the density of various molecules within molecular clouds. JWST has recently conducted extensive observations of molecular clouds, protoplanetary disks, and protostars \citep{2022PASP..134e4301B,2023NatAs...7..431M,2023A&A...673A.121B,2023ApJ...951L..32H,2024A&A...685A..73H,2024A&A...685A..74P,2024A&A...688A..26A,2024A&A...687A..86V,2024Sci...383..988B,2024A&A...687A..87F,2024ApJ...966...41F,2024ApJ...962L..16N}.

IRAS 16253-2429 is a nearby low-mass Class 0 protostar, characterized by its low temperature and a positive spectral index between $4.5 \, \mu m$ and $24 \, \mu m$ \citep{2023ApJS..266...32P}. This classification denotes the earliest phase of stellar evolution, where the protostar is still accreting mass and has not yet reached a stable configuration. As part of the Investigating Protostellar Accretion Program during JWST Cycle 1, IRAS 16253-2429 is recognized as the lowest luminosity source in a medium General Observer (GO) program \citep{2024ApJ...962L..16N}. It is situated at a distance of approximately 140 parsecs from Earth.
This protostellar system exhibits a bolometric luminosity of about $0.2\, \rm{L_{\odot}}$ and a temperature of $42 \, \rm{K}$ (see \cite{2024ApJ...962L..16N}, with further details forthcoming in Pokhrel et al., in preparation). The disk surrounding IRAS 16253-2429 is notably cold and is growing at a slow pace \citep{2019ApJ...871..100H}. Its characteristics have broadened our understanding of very low-mass star-forming processes, challenging the traditional paradigms \citep{2023ApJ...954..101A,2016ApJ...826...68H,2017ApJ...834..178Y}.
Recently, \cite{2024ApJ...962L..16N} reported identified a collimated jet associated with IRAS 16253-2429, featuring multiple [Fe II] emission lines alongside [Ne II], [Ni II], and H I emission lines, but notably absent in molecular emission. The spatial distribution and emission characteristics of this jet align with those reported in \cite{2024ApJ...966...41F}. The atomic jet emanating from IRAS 16253-2429 is oriented perpendicular to its cold protostellar disk and is observed to be moving at a velocity of $169 \pm 15\, \rm{km\, s^{-1}}$ \citep{2024ApJ...962L..16N}, providing valuable insights into the dynamics of early stellar evolution.

IRAS 23385+6053 is classified as a Class 0 protostar, characterized by a temperature of approximately $40 \, \rm{K}$ \citep{2004A&A...414..299F}. Despite its early evolutionary stage, this protostar is of high mass, estimated to be around $370 \, \rm{M_{\odot}}$ \citep{1998ApJ...505L..39M}, and exhibits a luminosity of approximately $3 \times 10^3 \, \rm{L_{\odot}}$ \citep{2019A&A...627A..68C}. This source represents a rapidly accreting, massive protostellar system that is in a very early phase of its development \citep{2012ApJ...745..116W}.
Previous observations, as reported in \cite{2012ApJ...745..116W}, have detected 95 GHz methanol ($\rm{CH_3OH}$) maser emission emanating from several locations within the core of this protostellar system. Additionally, \cite{2003A&A...407..237T} identified methanol emission lines near a frequency of $\sim338 \, \rm{GHz}$. More recent findings, as detailed in \cite{2024A&A...683A.249F}, include the detection of multi-phase water ($\rm{H_2O}$) features, $\rm{CO_2}$, and other spectral lines in the data collected by the JWST MIRI.

In this work, we estimate intensities of absorption of molecules $\rm{CH_4}$ ($\sim 7.7\, \mu m$), $\rm{CO_2}$ ($\sim 15\, \mu m$) , $\rm{H_2O}$  ($\sim 6\, \mu m$), $\rm{CH_3OH+NH_4^+}$ ($\sim 6.7\, \mu m$), silicate dust ($\sim 10\, \mu m$) and the emission line $\rm{[S\,I]}$ ($\sim 25.249\, \mu m$) of the two protostars, search for possible spatial distribution consistency among the absorption lines. 

\begin{figure*}[htbp]
\centering
\includegraphics[width=0.475\linewidth]{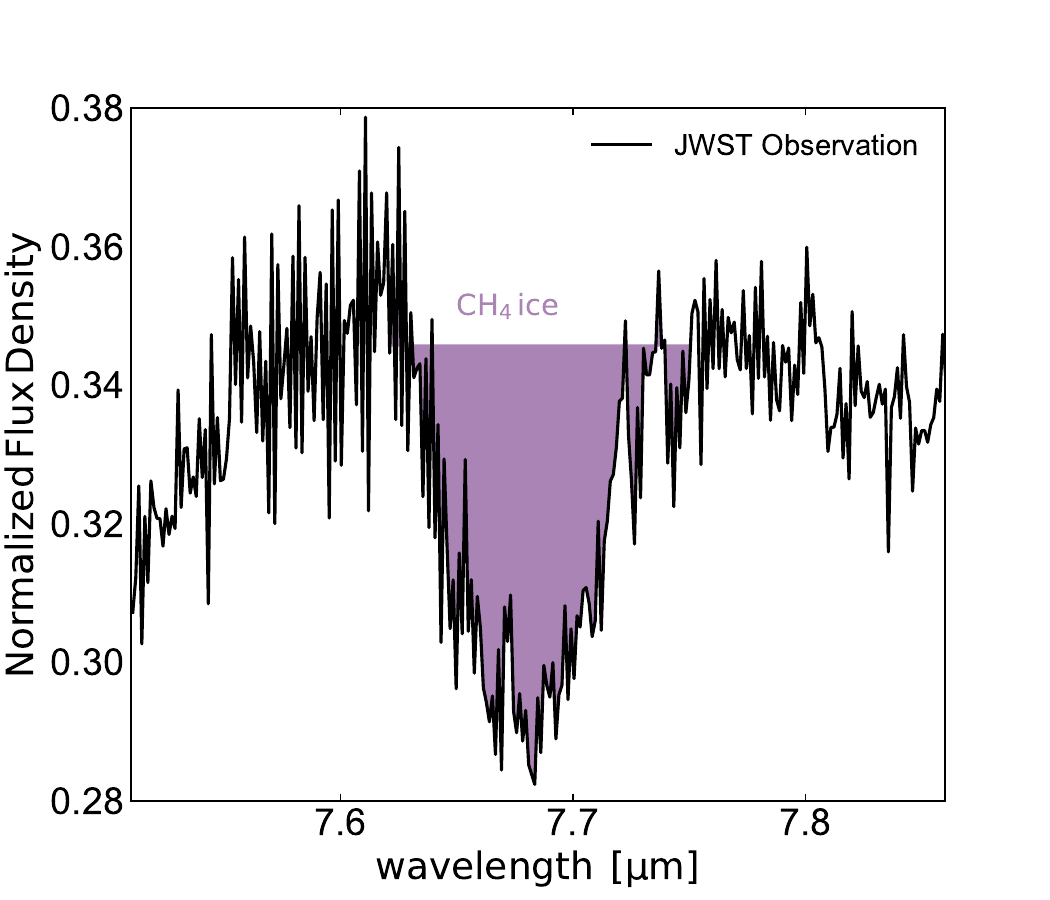}
\includegraphics[width=0.475\linewidth]{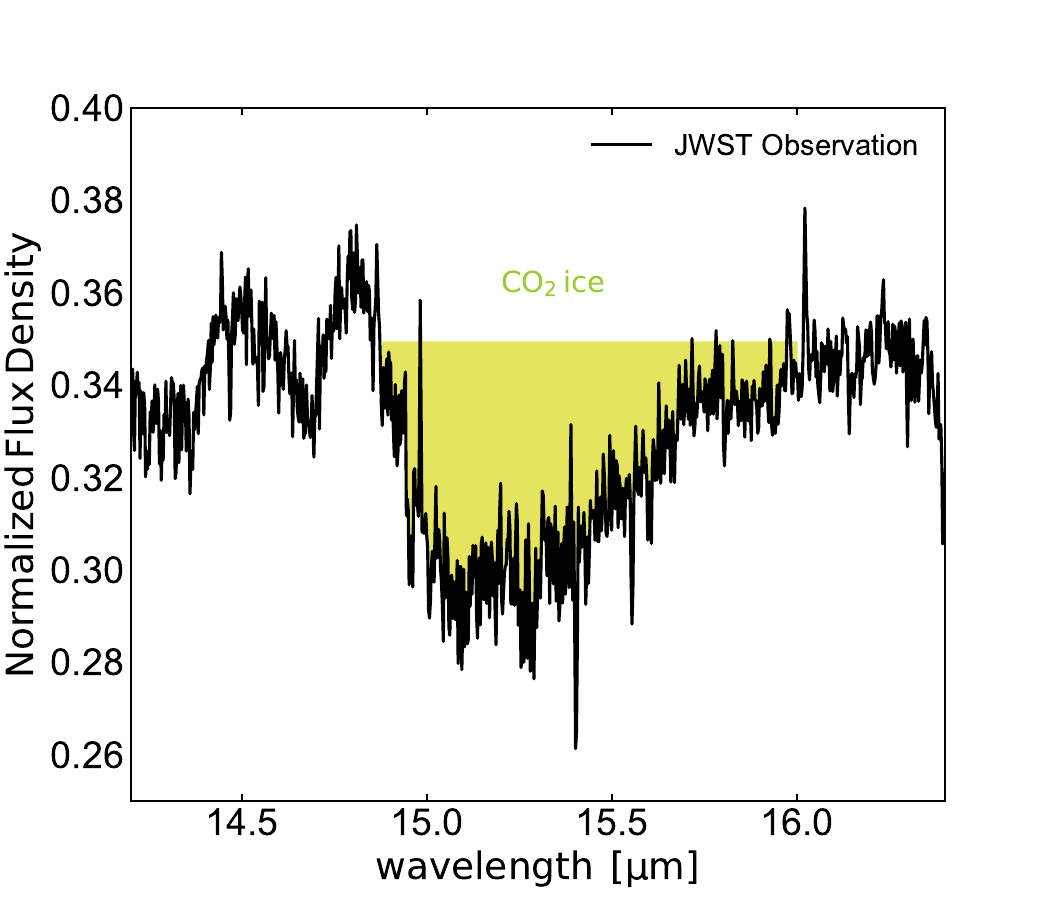}
\includegraphics[width=0.475\linewidth]{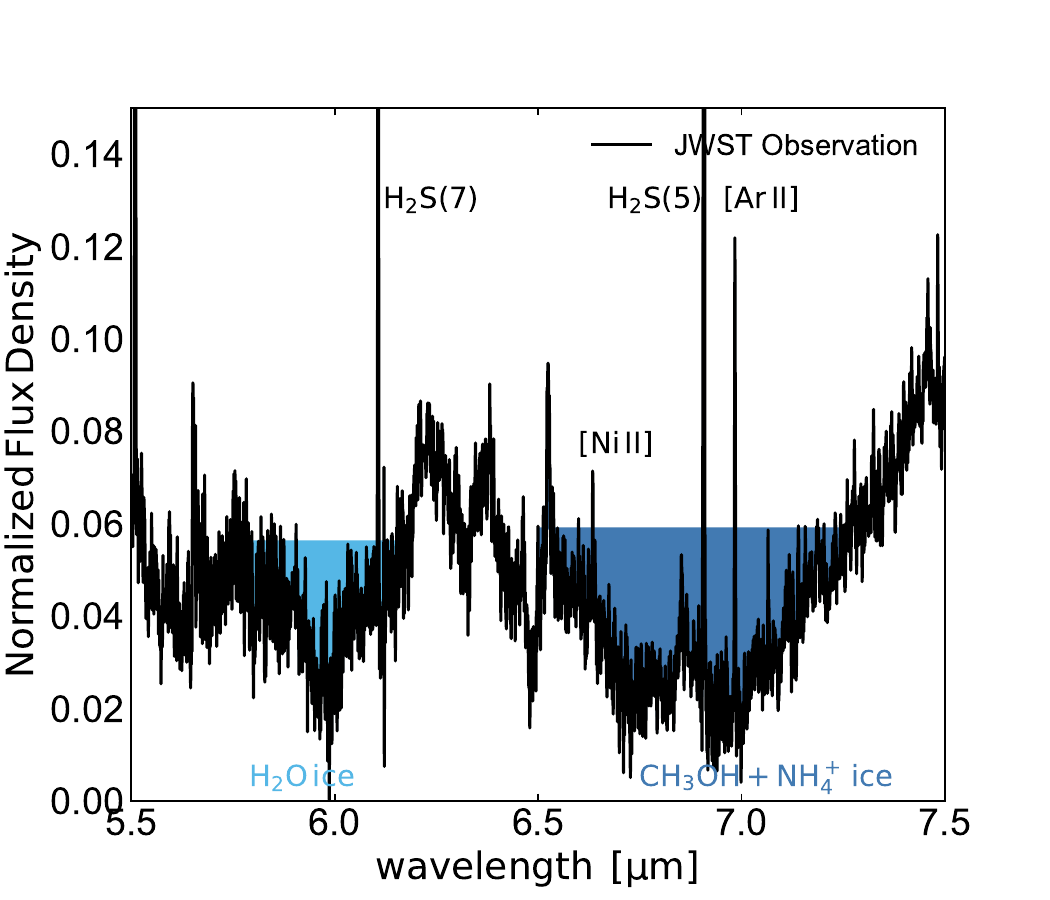}
\includegraphics[width=0.475\linewidth]{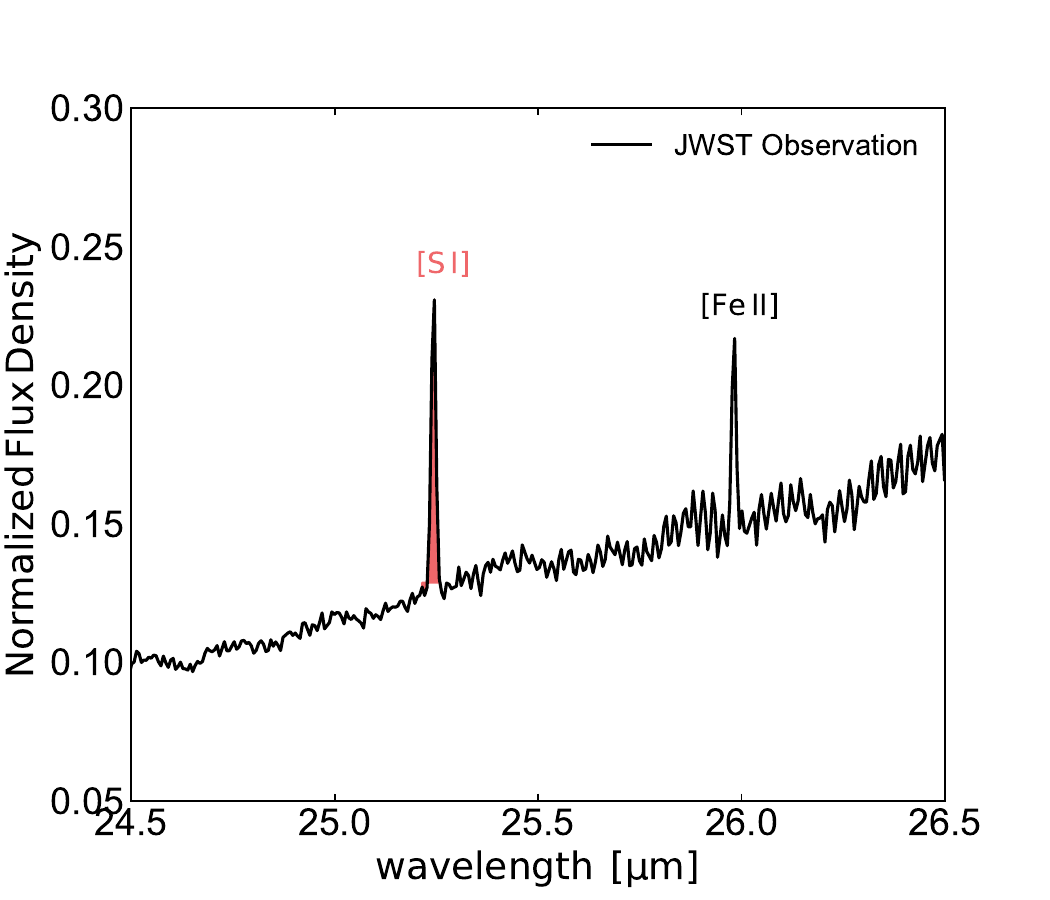}
\caption{\label{fig:1} The JWST observation of the ice absorption dips of $\rm{CH_4}$, $\rm{CO_2}$, $\rm{H_2O}$ , $\rm{CH_3OH+NH_4^+}$ molecules and the the emission line $\rm{[S\,I]}$ of the protostars. Upper Left Panel: The $\rm{CH_4}$ absorption dip feature in the stacked 1-dimension JWST/MIRI-IFU spectrum of protostar IRAS 23385+6053. The purple area marked the equivalent width of the $\rm{CH_4}$ absorption dip. Upper 
Right Panel: The $\rm{CO_2}$ absorption dip feature in the stacked 1-dimension JWST/MIRI-IFU spectrum of protostar IRAS 16253-2429. The purple area marked the equivalent width of the $\rm{CO_2}$ absorption dip. Bottom Left Panel: The ice absorption dips of $\rm{H_2O}$ and $\rm{CH_3OH+NH_4^+}$ in the stacked 1-dimension JWST/MIRI-IFU spectrum of protostar IRAS 23385+6053. The cyan and deep sky-blue marked the EWs of the two dips. Bottom Right Panel: The $\rm{[S\, I]}$ emission line in the stacked 1-dimension JWST/MIRI-IFU spectrum of protostar IRAS 23385+6053. The red area marked the equivalent width of the $\rm{[S\, I]}$ emission line.  }
\end{figure*}

\section{JWST Observations}
\label{JWST+Obs}
The protostellar object IRAS 16253-2429, located at right ascension (RA) 247.0926 degrees and declination (DEC) -24.6087 degrees, was observed by the JWST MIRI/IFU on July 23, 2022, under the direction of Principal Investigator (PI) Tom Megeath. The total observation time allocated for this task was approximately 7.4 hours.
Similarly, the protostellar object IRAS 23385+6053, with RA 355.2271 degrees and DEC 61.1744 degrees, was observed on August 22, 2022, also using the JWST MIRI/IFU. This observation was led by PI Ewine F. Van Dishoeck and had a total observation time of approximately 2.12 hours.

We retrieved the public JWST observation data for these two protostars from the Mikulski Archive for Space Telescopes (MAST) website\footnote{MAST:\url{https://mast.stsci.edu/}}. The data we obtained, classified as Level 3 Science data, have undergone standard reduction processes, which include the subtraction of background noise. These data are presented as three-dimensional cubes, with axes corresponding to spatial dimensions (Right Ascension and Declination) and wavelength, facilitating comprehensive analysis.

To determine the intensity of absorption or emission features within spectral data, we employ the calculation of equivalent width (EW) across appropriate wavelength intervals, as defined by the following equation:

\begin{equation}\label{eq:1}
	\rm{EW} = \mathit{\int_{\lambda_1}^{\lambda_2} \frac{ | f_{\nu,c}-f_{\nu,\lambda} | }{f_{\nu,c}} d \lambda} ,
\end{equation}

where $\lambda_1$ and $\lambda_2$ denote the lower and upper bounds of the wavelength range for the integration of the spectral feature in question. Here, $f_{\nu,c}$ represents the continuum spectral flux density, and $f_{\nu,\lambda}$ is the observed flux density of the absorption or emission line at a given wavelength $\lambda$. The value of $f_{\nu,c}$ is derived from the mean flux density in adjacent wavelength ranges that are free from significant spectral lines, thus representing a clean continuum spectrum.

For the purposes of this study, the continuum spectrum wavelength ranges selected are as follows: $7.8-7.9\, \mu m$ for $\rm{CH_4}$, $14.6-14.7\, \mu m$ for $\rm{CO_2}$, $6.15-6.20\, \mu m$ for $\rm{H_2O}$, $7.25-7.3\, \mu m$ for $\rm{CH_3OH+NH_4^+}$ and  $25.28-25.32\, \mu m$ for $\rm{[S\,I]}$. The integration wavelengths for the absorption and emission lines are: $7.55-7.7\, \mu m$ for $\rm{CH_4}$, $14.75-15.75\, \mu m$ for $\rm{CO_2}$, $5.76-6.15\, \mu m$ for $\rm{H_2O}$, $6.5-7.25\, \mu m$ for $\rm{CH_3OH+NH_4^+}$ and $25.21-25.26\, \mu m$ for $\rm{[S\, I]}$. The silicate absorption feature is nearly fully saturated, leading to a significant extinction of the underlying continuum emission, as reported in \cite{2023A&A...673A.121B}. Consequently, we employ the differential continuum emission between 7.8 micrometers and 10.65 micrometers as a proxy for the equivalent width (EW) to gauge the intensity of the silicate absorption feature.

To ensure the reliability of our analysis and to mitigate the impact of extreme hot pixels, we have established criteria for the exclusion of data points during the integration process. Specifically, any data point with a flux density $f < 10^{-3} \, \rm{mJy\, sr^{-1}}$ or $f > 10^{3} \, \rm{mJy\, sr^{-1}}$ is omitted from the integration for each pixel. This approach helps to maintain the integrity of the spectral analysis and to reduce potential errors associated with outlier values.

For absorption dip from molecules, the EW is $	\rm{EW_{abs}} = \mathit{\int_{\lambda_1}^{\lambda_2} \frac{f_{\nu,c}-f_{\nu,\lambda}}{f_{\nu,c}} d \lambda}$. The background subtracting of data reduction process with standard pipeline will lead to negative flux density in some pixels. But that is not a important problem for this work, because we estimate the relativistic intensity of absorption or emission lines with Equation.~(\ref{eq:1}).

Figure~\ref{fig:1} shows the JWST data of the  absorption dips of $\rm{CH_4}$, $\rm{CO_2}$, $\rm{H_2O}$, $\rm{CH_3OH+NH_4^+}$ molecules and the silicate absorption of the protostars. The upper left panel displays the $\rm{CH_4}$ absorption dip in the 1-dimensional stacked spectrum of protostar IRAS 23385+6053, as captured by the JWST/MIRI-IFU. The purple shaded region indicates the equivalent width (EW) of the $\rm{CH_4}$ absorption dip, providing a quantitative measure of this spectral feature. The upper right panel illustrates the $\rm{CO_2}$ absorption dip in the 1-dimensional stacked spectrum of protostar IRAS 16253-2429. Similar to the left panel, the purple shaded region represents the EW of the $\rm{CO_2}$ absorption dip, offering a comparable quantitative assessment. The bottom left panel shows the absorption features of $\rm{H_2O}$ and $\rm{CH_3OH+NH_4^+}$ in the 1-dimensional stacked spectrum of protostar IRAS 23385+6053.  The bottom left panel showcases the $\rm{[S\,I]}$ emission line in the IRAS 23385+6053 spectrum.

\begin{figure*}[htbp]
\centering
\includegraphics[width=0.32\linewidth]{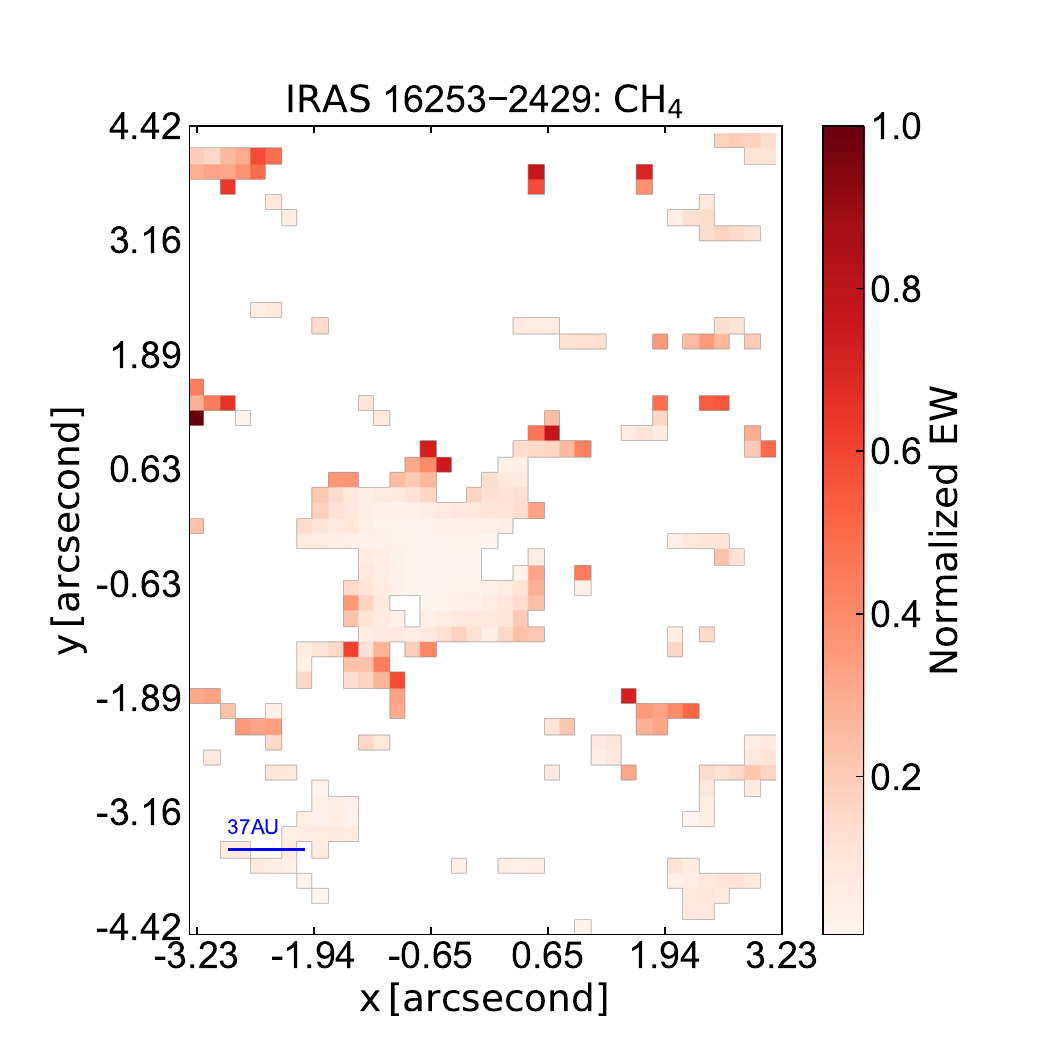}
\includegraphics[width=0.32\linewidth]{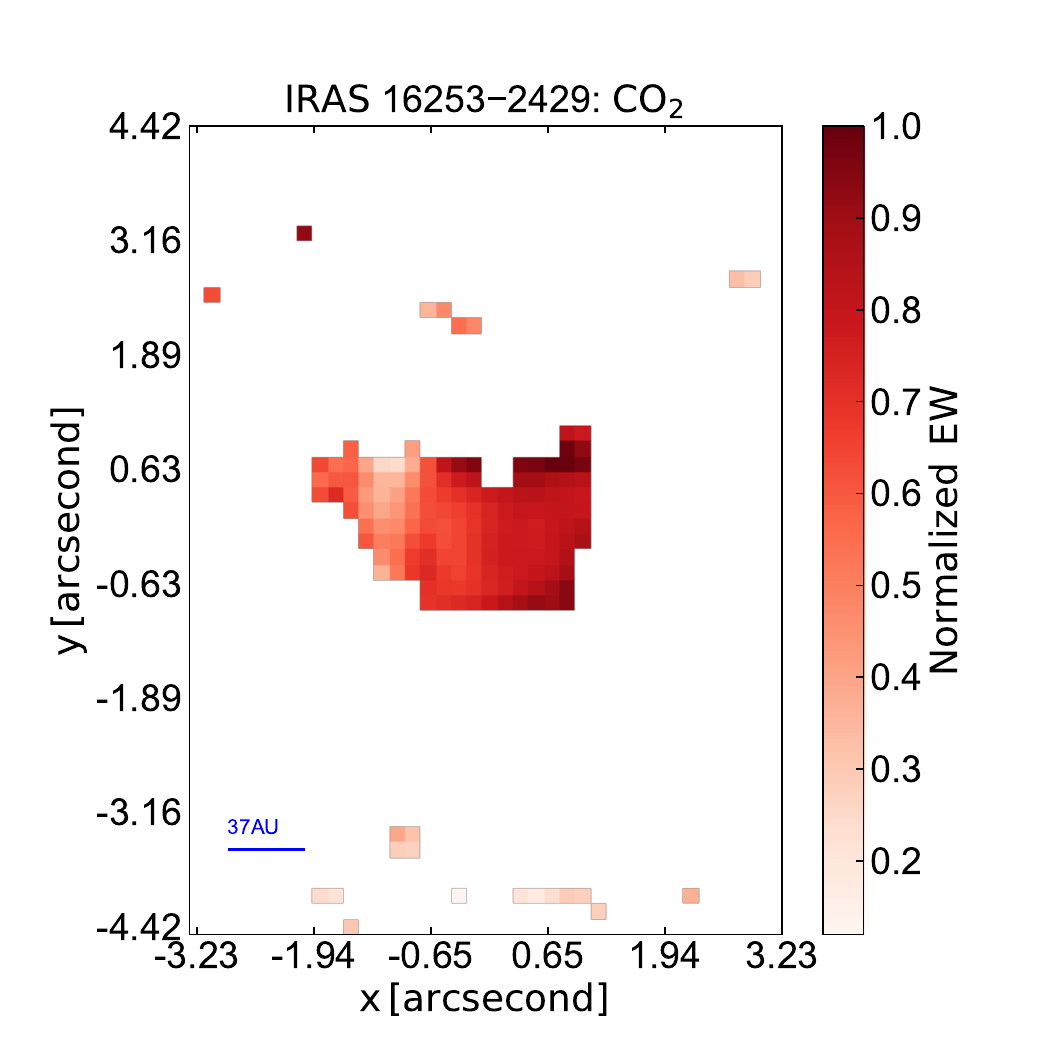}
\includegraphics[width=0.32\linewidth]{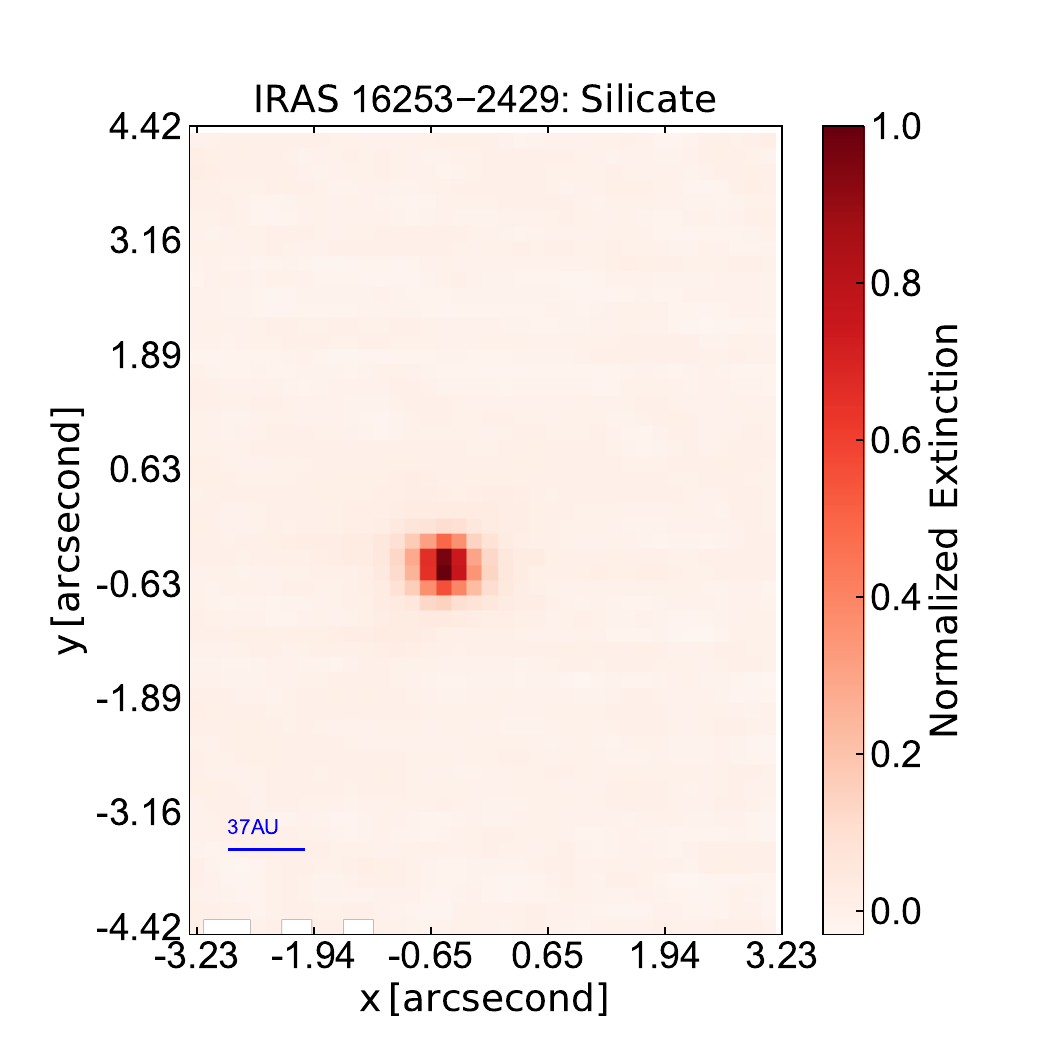}
\includegraphics[width=0.32\linewidth]{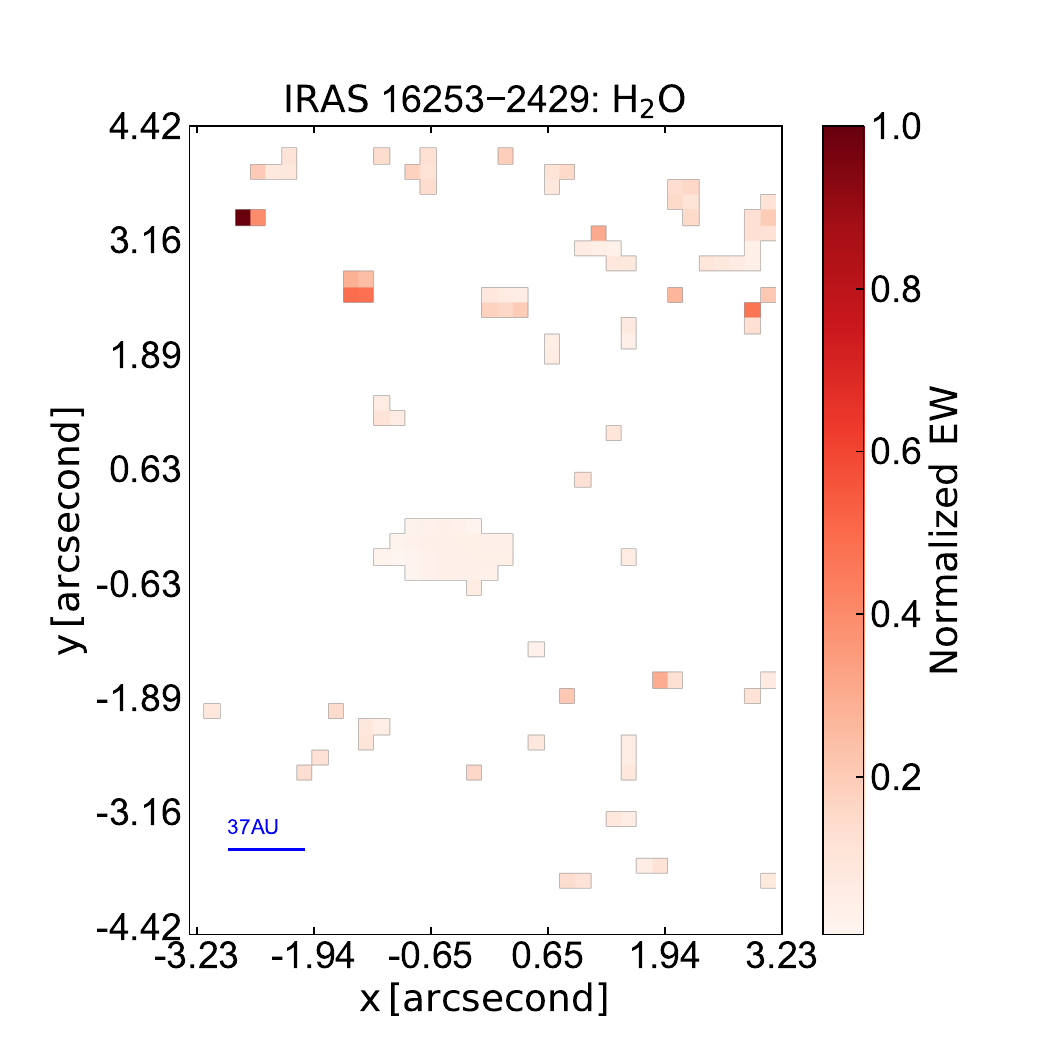}
\includegraphics[width=0.32\linewidth]{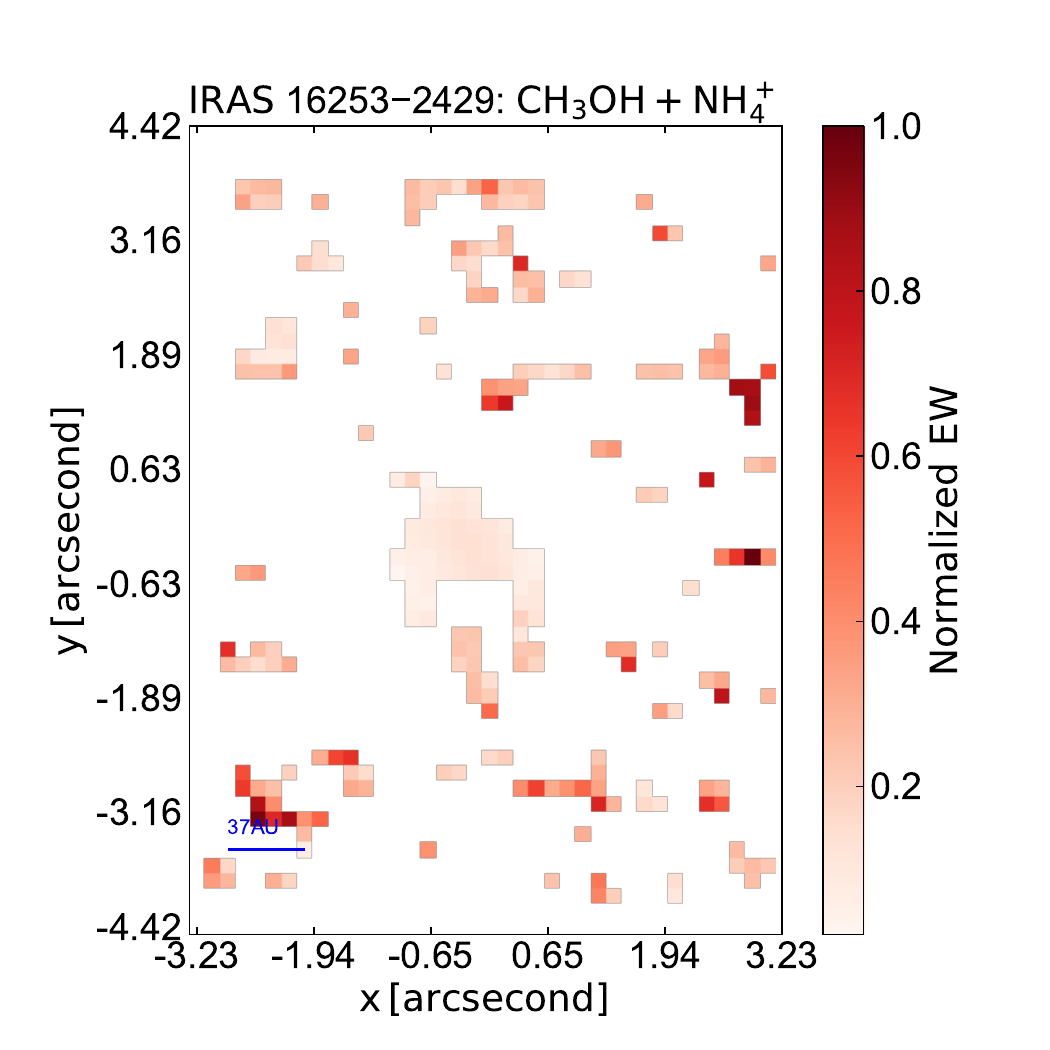}
\includegraphics[width=0.32\linewidth]{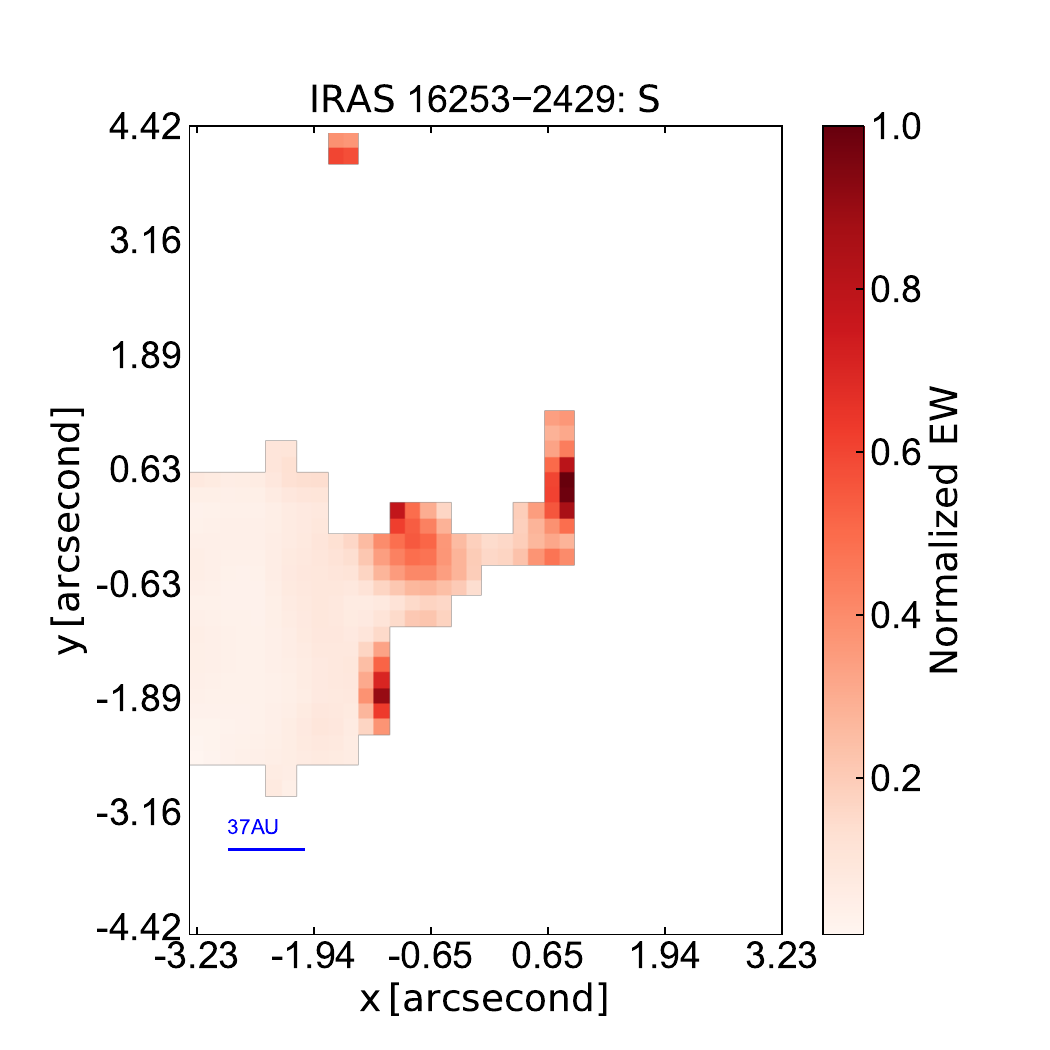}
\includegraphics[width=0.32\linewidth]{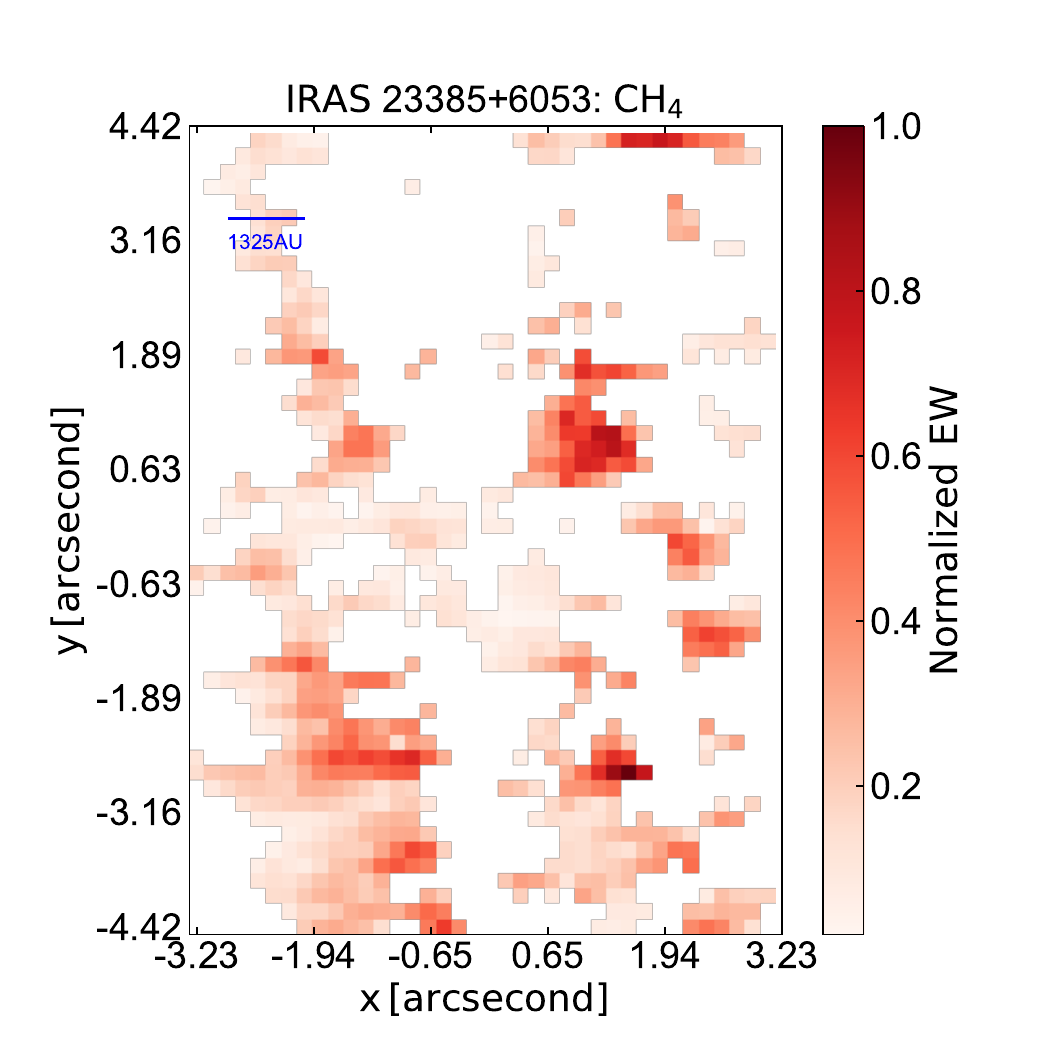}
\includegraphics[width=0.32\linewidth]{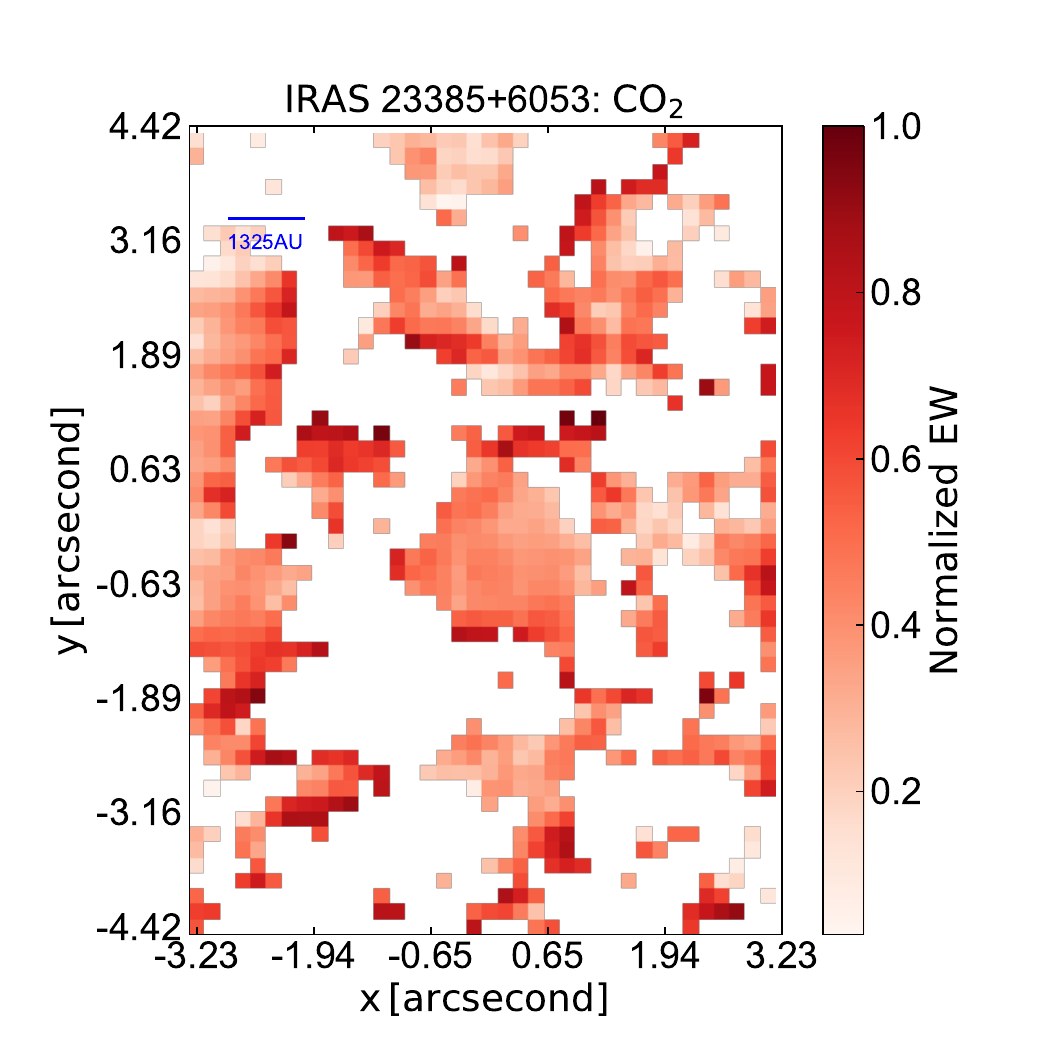}
\includegraphics[width=0.32\linewidth]{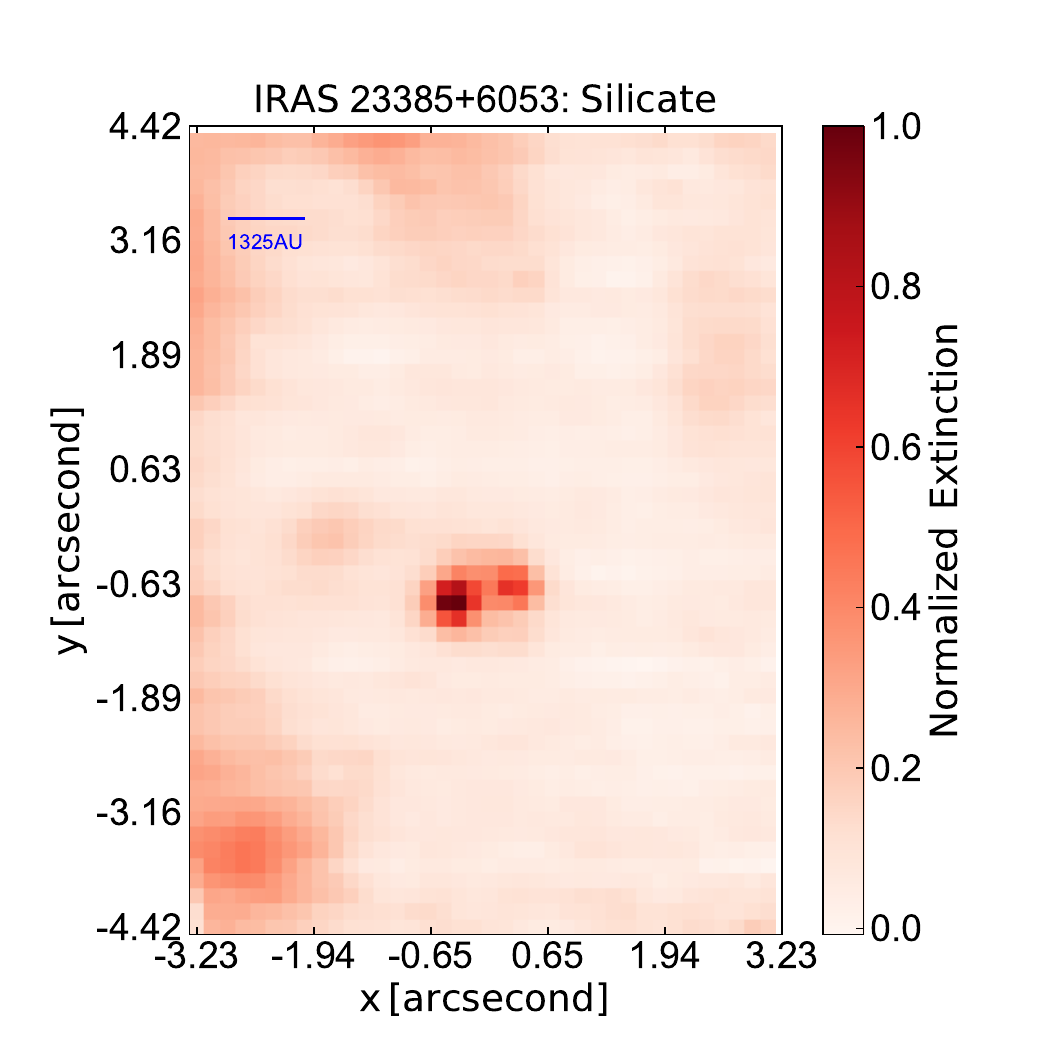}
\includegraphics[width=0.32\linewidth]{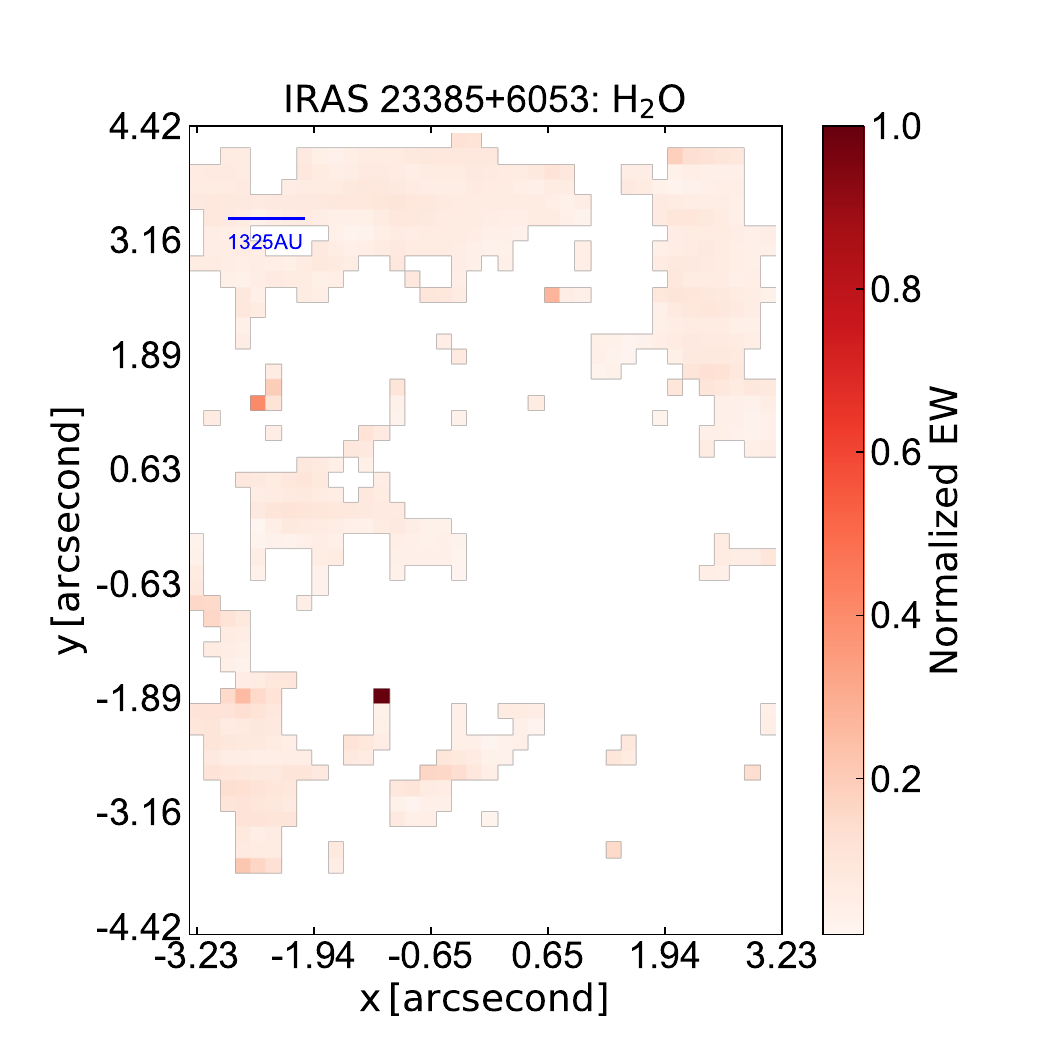}
\includegraphics[width=0.32\linewidth]{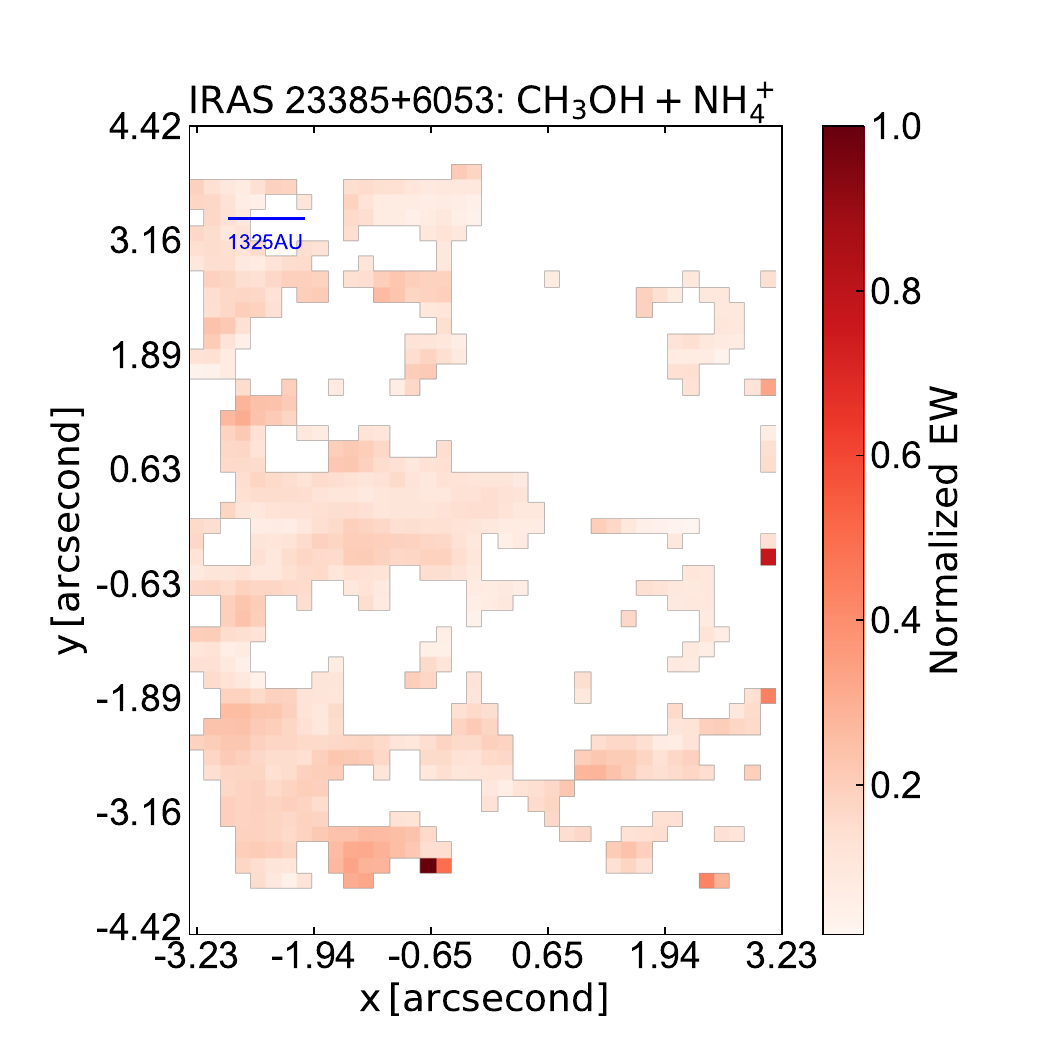}
\includegraphics[width=0.32\linewidth]{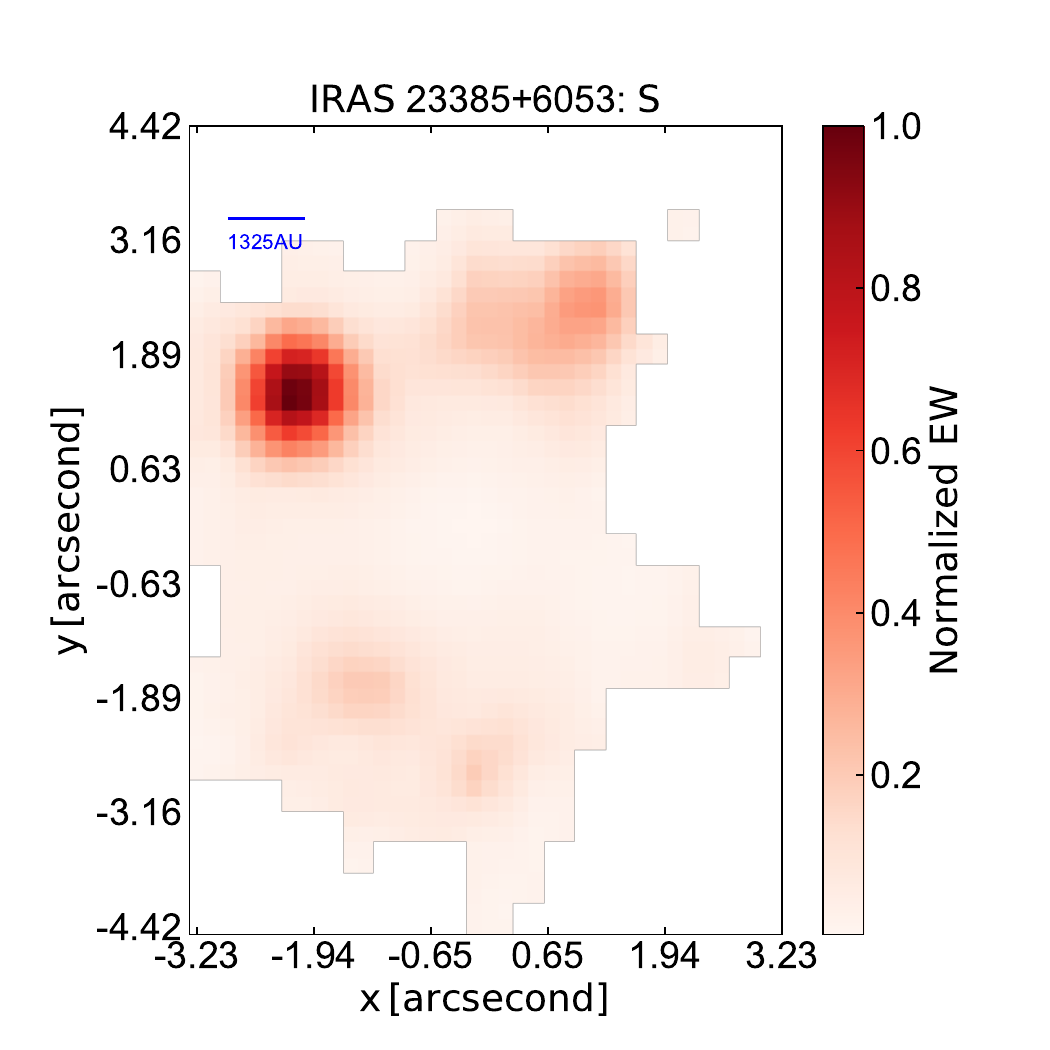}
\caption{\label{fig:2} The equivalent width of the  absorption dips of $\rm{CH_4}$, $\rm{CO_2}$, $\rm{H_2O}$ , $\rm{CH_3OH+NH_4^+}$ molecules, silicate extinction and the $\rm{[S\, I]}$ atoms emission line from JWST MIRI/IFU observation. Top Two Panels: The results of the protostar IRAS 16253-2429. Bottom Two Panels: The results of the protostar IRAS 23385+6053. The liner scale-bar with a size of 0.85 arcsecond is in blue. }
\end{figure*}

\section{Spatial Distribution of Molecules and Dust}

After determining the molecular absorption equivalent width and silicate extinction values for the protostars IRAS 16253-2429 and IRAS 23385+6053, we proceeded to enhance the data quality by manipulating the EW maps using the OpenCV package for Python. This step was necessary due to the inferior data quality at the four corners of the original data cube.

The James Webb Space Telescope's Mid-Infrared Instrument (MIRI) Integral Field Unit (IFU) comprises four channels, designated as ch1, ch2, ch3, and ch4, each with distinct fields of view. The absorption dips of $\rm{H_2O}$ at $\sim 6\mu m$ and $\rm{CH_3OH+NH_4^+}$ at $\sim 6.7\mu m$ were stored in the ch2 data cube. The absorption features of $\rm{CH_4}$ at approximately $7.7 \, \mu m$ and silicate at approximately $10 \, \mu m$  were captured in the ch2 data cube, while the $\rm{CO_2}$ absorptions at approximately $15 \, \mu m$ were observed in the ch3 data cube. The emission line $\rm{[S\,I]}$ ($\sim 25.249\, \mu m$) is stored in ch4 data cube.

Given these observations, we utilized OpenCV to perform the following steps:
1. Rotation and Cropping: We rotated each map to align with the data quality parameters and cropped out a square area that exhibited good data quality, avoiding the poor quality corners.
2. Zooming: We zoomed out on the central regions of the $\rm{H_2O}$ and $\rm{CH_3OH+NH_4^+}$ absorption maps. The zoom scale was adjusted to match the field of view ratios of ch1 to ch2. We zoomed in on the central regions of the $\rm{CO_2}$ absorption maps. The zoom scale was adjusted to match the field of view ratios of ch3 to ch2 for the $\rm{CO_2}$ maps. That scale was adjusted to match the field of view ratios of ch4 to ch2 for the $\rm{[S\, I]}$ maps.
3. Uniform Scaling: To ensure consistency across the maps, we resized the maps to a uniform scale of approximately $6.5\, \rm{arcseconds} \times 9.0\, \rm{arcseconds}$.

This meticulous processing allowed us to generate maps that are not only of higher quality but also comparable in scale, facilitating a more accurate analysis and comparison of the spectral features across the different molecular and atomic species observed in the protostellar environments.
The final maps provide a clearer representation of the absorption and emission features.

Figure~\ref{fig:2} presents an analysis of the equivalent widths of the absorption dips for $\rm{CH_4}$, $\rm{H_2O}$, $\rm{CH_3OH+NH_4^+}$, $\rm{CO_2}$, silicate extinctions and equivalent widths of sulfur emission lines, as observed by the JWST MIRI/IFU. The visualizations are standardized to the same field of view for a direct comparison of the spatial distributions of these species across different protostellar environments.The EW values are normalized, allowing for a comparison of relative intensities across the maps.

The top two panels depict the maps for the protostar IRAS 16253-2429.
The $\rm{CH_4}$, $\rm{H_2O}$, $\rm{CH_3OH+NH_4^+}$ and $\rm{CO_2}$ molecules exhibit a consistent spatial distribution in the central region of the protostar, suggesting a potential correlation between these species.The distribution of silicate dust in the protostar IRAS 16253-2429 is concentrated to the protostellar disk. The jet/outflow powered sulfur atoms map is much different to these absorption dips. 

The bottom two panels showcase the maps for the high-mass protostar IRAS 23385+6053.
The morphologies of the molecules and silicate dust for this protostar display distinct patterns, suggesting diverse physical and chemical processes at play.
For IRAS 23385+6053, the result highlighting the complexity and variability of protostellar environments.

\begin{table}[t!]
\caption{\label{tab:1}%
The structural similarity (SSIM) of the molecules and dust maps.
}
\begin{tabular}{ccc}
\hline 
\textrm{Maps} &
\textrm{SSIM} &
\textrm{SSIM} \\
\textrm{ } &
\textrm{IRAS} &
\textrm{IRAS} \\
\textrm{ } &
\textrm{16253-2429} &
\textrm{23385+6053} \\
\hline
$\rm{CH_4}$ \& $\rm{CO_2}$ & 0.55 & 0.07 \\ 
$\rm{CH_4}$ \& $\rm{H_2O}$ & 0.55 & 0.26 \\ 
$\rm{CH_4}$ \& $\rm{[S\, I]}$ & 0.23 & 0.24 \\ 
$\rm{CH_4}$ \& $\rm{Silicate}$ & 0.51 & 0.15 \\ 
$\rm{CH_4}$ \& $\rm{Silicate+CO_2}$ & 0.46 & 0.23 \\ 
$\rm{CH_4}$ \& $\rm{CH_3OH+NH_4^+}$ & 0.35 & 0.22 \\ 
\hline
\end{tabular}
\end{table}

We employed the concept of structural similarity (SSIM) to quantitatively assess the degree of similarity between different molecular maps and the dust map for the protostars IRAS 16253-2429 and IRAS 23385+6053. The SSIM is based on the work of \cite{1284395,1415469,4775883}. Our calculations, as detailed in Table~\ref{tab:1}, reveal that the SSIM values among the $\rm{CH_4}$, $\rm{CO_2}$, and $\rm{H_2O}$ maps for IRAS 16253-2429 are notably high. 
Conversely, the SSIM between the $\rm{CH_4}$ map and the jet/outflow powered sulfur emission line map is found to be very low. This discrepancy indicates a difference in the spatial distribution patterns of these species. The high SSIM values among $\rm{CH_4}$, $\rm{CO_2}$ and $\rm{H_2O}$ maps hint at a coherent molecular distribution, which could be indicative of a shared origin or interaction.  This observation is in line with the "Classical" dark-cloud chemistry model, which emphasizes the role of ion-molecule reactions in the chemical evolution of these environments \citep{1973ApJ...185..505H}. On the other hand, the low SSIM between $\rm{CH_4}$ and sulfur atoms maps suggests a weaker spatial correlation, which may point towards different mechanisms governing their distribution.

\section{Conclusion and Discussion}
\label{conclusion}

In our research, we have conducted a comprehensive analysis to estimate the intensities of molecular absorption for $\rm{CH_4}$ (at approximately $7.7\, \mu m$), $\rm{H_2O}$ (at approximately $6\, \mu m$), $\rm{CH_3OH+NH_4^+}$ (at approximately $6.7\, \mu m$),  $\rm{CO_2}$ (at approximately $15.0\, \mu m$),  silicate dust (at approximately $10.0\, \mu m$), and the emission from sulfur atoms $\rm{[S\, I]}$ (at approximately $25.249\, \mu m$) in two protostellar sources. Our primary objective was to investigate the spatial distribution consistency among $\rm{CH_4}$, $\rm{H_2O}$ and $\rm{CO_2}$.

The high SSIM value between $\rm CH_4$ and $\rm CO_2$ maps due to surrounded distribution in IRAS 16253-2429 hints that carbon dioxide may play an important role in the production of methane. Hydrogen molecules adsorbed on the surface of solid carbon dioxide may undergo hydrogenation reactions under the action of cosmic rays to form methane and water molecules. ALthough the distribution of the simple molecules in the protostar IRAS 16253-2429 is consistent with the "Classical" dark-cloud chemistry with ion-molecule reactions \citep{1973ApJ...185..505H,2009ARA&A..47..427H,2012A&ARv..20...56C}, methanogenic life attached to the surface of solid carbon dioxide may also undergo the aforementioned biochemical processes, as predicted in \cite{2023arXiv231114291F}. 
These results highlight the importance of continued astrobiological studies and the need for advanced observational techniques to unravel the complex chemistry of protostellar systems.

Our analysis of the spatial distribution of $\rm{CH_4}$, $\rm{CO_2}$, and $\rm{H_2O}$ in the protostellar system IRAS 16253-2429 indicates a high degree of spatial consistency among these molecules. This observation suggests that complex organic molecules, such as $\rm CH_4$, are likely formed and undergo freeze-out processes in the early stages of protostellar envelope evolution. However, these molecules may be susceptible to destruction by the intense shocks associated with outflows or accretion activities during star formation, as evidenced by the irregular and inconsistent distribution observed in the maps of IRAS 23385+6053.

ALMA observations have indeed provided valuable insights into the protostellar disk of IRAS 16253-2429, suggesting a very low growth rate \citep{2019ApJ...871..100H}. This characteristic indicates a significant divergence in the evolutionary paths of low-mass protostars as compared to their high-mass counterparts. The high-mass protostar IRAS 23385+6053 may exhibit instability due to its accretion and turbulence processes \citep{1998ApJ...505L..39M,2004A&A...414..299F,2012ApJ...745..116W,2024A&A...683A.249F}, which could have profound implications for its developmental trajectory.

The future employment of multi-band observations, leveraging the capabilities of both ALMA and JWST, holds great promise for advancing our understanding of these protostellar systems. By combining the high-resolution imaging and spectroscopic prowess of ALMA with the exquisite sensitivity and infrared capabilities of JWST, we can expect to gain a more comprehensive view of the physical and chemical processes at play in these early stellar environments.

\begin{acknowledgements}
      We thank Qiao Li and Fu-Jun Du for the helpful discussion. This work is based on observations made with the NASA/ESA/CSA James Webb Space Telescope. The data were obtained from the Mikulski Archive for Space Telescopes at the Space Telescope Science Institute, which is operated by the Association of Universities for Research in Astronomy, Inc., under NASA contract NAS 5-03127 for JWST. These observations are associated with the program JOYS (JWST Observations of Young protoStars, proposal id. 1290 \& 1802).  
\end{acknowledgements}

%

\bibliographystyle{aa} 
\bibliography{biblio} 

%


%
%
\end{document}